\pdfoutput=1
\documentclass[aps,twocolumn,showpacs,floatfix,superscriptaddress]{revtex4-1}
\usepackage{graphicx}
\usepackage{amssymb}
\usepackage{color}
\usepackage{amsmath}
\usepackage{ulem}
\usepackage{mathrsfs}
\newcommand{\be}{\begin{equation}}
\newcommand{\ee}{\end{equation}}
\newcommand{\la}{\langle}
\newcommand{\ra}{\rangle}

\newcommand{\mb}{\mathbf}

\begin{document}

\title{Level spectroscopy in a two-dimensional quantum magnet: \\ Linearly dispersing spinons at the deconfined quantum critical point} 
  
\author{Hidemaro Suwa}
\affiliation{Department of Physics, University of Tokyo, Tokyo 113-0033, Japan}
\affiliation{Department of Physics, Boston University, 590 Commonwealth Avenue, Boston, Massachusetts 02215, USA}

\author{Arnab Sen} 
\affiliation{Department of Theoretical Physics, Indian Association for the Cultivation of Science, Jadavpur, Kolkata 700032, India}

\author{Anders W. Sandvik}
\affiliation{Department of Physics, Boston University, 590 Commonwealth Avenue, Boston, Massachusetts 02215, USA}

\begin{abstract}
  We study the level structure of excitations at the ``deconfined'' critical point separating antiferromagnetic
  and valence-bond-solid phases in two-dimensional quantum spin systems using the $J$-$Q$ model as an example. Energy gaps
  in different spin ($S$) and momentum (${\bf k}$) sectors are extracted from imaginary-time correlation functions obtained
  in quantum Monte Carlo simulations. We find strong quantitative evidence for
  deconfined linearly dispersing spinons with gapless points at ${\bf k}=(0,0)$, $(\pi,0)$, $(0,\pi)$,
  and $(\pi,\pi)$, as inferred from two-spinon excitations ($S=0$ and $S=1$ states) around these points. We also
  observe a duality between singlet and triplet excitations at the critical point and inside the ordered phases,
  in support of an enhanced symmetry, possibly SO(5). 
\end{abstract}

\date{\today}

\pacs{75.10.Jm, 75.40.Mg, 75.30.Ds, 75.40.Cx}

\maketitle

\section{Introduction}

Conventional quantum phase transitions between different ground states of quantum many-body systems 
can be understood within the Landau-Ginzburg-Wilson (LGW) paradigm, according to which a critical point is
described by an order parameter whose fluctuation diverges\,\cite{Fisher1989,ChubukovSY1994}. Following intriguing
numerical results pointing to violations of LGW predictions \cite{Sandvik2002,Motrunich2004}, the deconfined
quantum critical (DQC) point was proposed as a scenario beyond the standard paradigm\,\cite{SenthilVBSF2004,SenthilBSVF2004}.
Here the low-energy physics is not described directly by order parameters, but by fractional degrees of freedom that emerge (deconfine)
on long length scales close to the DQC point. These fractional objects should have prominent signatures in excitation spectra and
experimentally accessible spectral functions. We here present a numerical study of low-energy
excitations at the DQC point of a two-dimensional (2D) quantum magnet.

The DQC point considered here separates  states with N\'eel antiferromagnetic (AFM) order and spontaneous dimerization
(valence-bond-solid, VBS, order)\,\cite{ReadS1990}, realized with the $J$-$Q$ spin-$1/2$ Hamiltonian\,\cite{Sandvik2007} 
\begin{equation}
  H=-J \sum_{\la ij \ra} P_{ij} - Q \sum_{\la ijkl \ra} P_{ij}P_{kl},
  \label{H_jq}
\end{equation}
where $P_{ij}=1/4 - {\bf S}_i \cdot {\bf S}_j$ is a singlet projector on sites $ij$, $\la ij\ra$ denotes
nearest-neighbor sites (links) on a periodic square lattice with $L^2$ sites, and $\la ijkl \ra$
denotes a pair of links on a $2\times 2$ site plaquette. The summations are over all links and plaquettes; thus  $H$ maintains
all the symmetries of the square lattice. The $Q=0$ case is the standard AFM-ordered Heisenberg model \cite{Manousakis1990}, and
when $Q/J$ is sufficiently large, $Q/J \agt 22$, projection of correlated singlets leads to columnar dimerization and loss of AFM order.
In contrast to frustrated Heisenberg systems that may also harbor VBS states and DQC points \cite{Wang2011,LiBHS2012,GongZSMF2014},
the $J$-$Q$ model is not affected by sign problems and can be studied using quantum Monte Carlo (QMC) simulations on large
lattices \cite{Sandvik2010b}.

The existence of the DQC point has been addressed in numerous studies of the $J$-$Q$ model
\,\cite{Sandvik2007,MelkoK2008,JiangNCW2008,Sandvik2010a,Sandvik2010b,Kaul2011,HaradaSOMLWTK2013,ChenHDKPS2013,BlockMK2013,PujariAD2015},
3D close-packed loop\,\cite{NahumSCOS2015} and dimer\,\cite{SreejithP2014} models (which provide effective descriptions of quantum spins),
and lattice versions of the proposed \,\cite{SenthilVBSF2004,SenthilBSVF2004} non-compact CP$^1$ DQC field
theory\,\cite{ChenHDKPS2013,KuklovMPST2008,MotrunichV2008}. Unusual scaling behaviors  were observed in these studies that were
not predicted within the DQC theory but which can now be accounted for by a scaling
hypothesis incorporating the two divergent length scales of the theory, a standard correlation length and a scale related to emergent U(1)
symmetry of the VBS fluctuations \cite{ShaoGS2016}. While there are still important unsettled questions remaining, e.g., on the fundamental
origins of the anomalous scaling \cite{NahumSCOS2015,ShaoGS2016} and an apparent emergent SO(5) symmetry \cite{Nahum2015b},
there is now little doubt that the transition is continuous (instead of weakly first order, as had been claimed in some studies
\cite{JiangNCW2008,KuklovMPST2008,ChenHDKPS2013}).

Dynamical properties of DQC systems have not been addressed in direct numerical calculations. The $J$-$Q$ model offers
unique opportunities to study deconfined excitations and the quantum dynamics of confinement. The deconfined excitations should
be spinons carrying spin $S=1/2$\,\cite{SenthilVBSF2004,SenthilBSVF2004}. Going into the ordered phases, pairs of spinons become
confined (bound) into $S=1$ magnons which are gapped in the VBS phase and gapless in the AFM phase. The existence of spinons has been
inferred from studies of $S=1$ states in QMC simulations \cite{Tang2013,ShaoGS2016}. However, the spinon dispersion relation has not
been computed and it has not been directly confirmed that the lowest singlets and triplets are degenerate, as they should be in
an infinite lattice with two independently propagating spinons. This degeneracy may not even be perfect, due to weak (logarithmic)
interactions between vortexlike spinons \cite{SenthilBSVF2004}.

Here we report QMC studies of the level spectrum of the $J$-$Q$ model at its DQC point. We analyze gaps extracted
from correlation functions, thus characterizing the level spectrum of spinons and scaling behaviors as bound states (magnons) 
form in the ordered phases. Our study reveals gapless critical $S=0$ and $S=1$ excitations at ${\bf k}=(0,0),(\pi,0),(0,\pi)$, and $(\pi,\pi)$,
and all these points are characterized by linear dispersion with a common velocity, thus lending strong support to elementary
$S=1/2$ spinons with dispersion minimums at the above four ${\bf k}$-points. Moreover, the scaling of singlet and triplet gaps in
the ordered phases exhibits a duality consistent with emergent SO(5) symmetry \cite{Senthil2006,Nahum2015b}.

The outline of the rest of the paper is as follows: In Sec.~\ref{sec:gapmethod} we explain the technical details of extracting gaps
from imaginary-time correlation functions. The scaling procedures used in combination with level spectroscopy based on finite-size 
gaps are presented in Sec.~\ref{sec:spectroscopy} along with results. In Sec.~\ref{sec:lineardisp}, the full dispersion relation along
a path in the Brillouin zone is discussed first, before a detailed analysis of linearly dispersing spinons in the neighborhood of the 
four gapless points. We briefly summarize our study and discuss implications in Sec.~\ref{sec:summary}.

\begin{figure}
  \center{\includegraphics[width=8cm, clip]{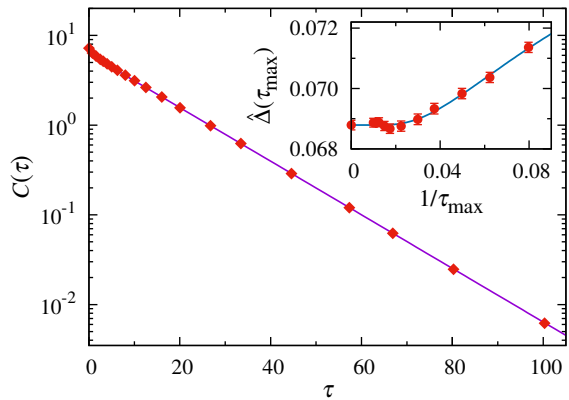}}
\vskip-2mm
\caption{(Color online) Gap estimation for $J/Q=0.045$, $L=64$, $S=1$, and ${\bf k}=(\pi,\pi)$. The main panel shows the imaginary-time 
dynamic correlation function. The gap extrapolation, resulting in $\hat{\Delta}(\infty)=0.06879(16)$, is illustrated in the inset.
The curve shows the fitting function $\hat{\Delta}(\tau_{\rm max}) - \hat{\Delta}(\infty) \propto e^{ - a \tau_{\rm max} }$, where 
the parameter $a$ is optimized for the best fit \,\cite{SuwaT2015,SenSS2015}. The straight line in the main panel has slope corresponding 
to the extracted gap (the prefactor being the sole fitting parameter).}
\label{fig:gap-pipi}
\vskip-2mm
\end{figure}

\begin{figure}
  \center{\includegraphics[width=8cm, clip]{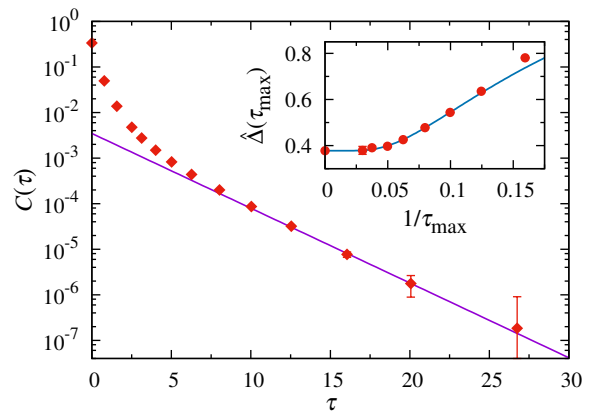}}
\vskip-2mm
\caption{(Color online) Gap estimation for $J/Q=0.045$, $L=64$, $S=1$, and ${\bf k}=(\pi,0)$, the procedures and fitted functions 
are analogous to those explained in Fig.~\ref{fig:gap-pipi}. The estimated gap is $\hat{\Delta}(\infty)=0.3778(36)$.}
\label{fig:gap-pi0}
\vskip-2mm
\end{figure}

\section{Gap Estimation by QMC}
\label{sec:gapmethod}

We have used finite-temperature and projector QMC simulations, with continuous-time worldlines as well as the
stochastic series expansion of ${\rm e}^{-\beta H}$ (where $\beta$ is either the inverse temperature or the projection
``time'') and sampling with loop updates\,\cite{Sandvik2010b,SandvikE2010}. In the projector approach we use a singlet-sector amplitude-product
trial state\,\cite{LiangDA1988,SandvikB2007}. All the methods gave mutually consistent results for sufficiently large $\beta$.
Most of the results reported here were obtained with the somewhat more efficient projector method.

To compute gaps for given $S$ and ${\bf k}$, an operator is chosen which transfers the quantum numbers upon exciting the
$S=0$, ${\bf k}=(0,0)$ ground state. The following operators were used for triplets and singlets, respectively:
\begin{equation}
  \mathscr{T}_{\mb k} =\sum_{\mb r} S^z_{\mb r} {\rm e}^{i {\mb r} \cdot {\mb k}},~~~
  \mathscr{S}_{\mb k}=\sum_{\mb r} S^z_{\mb r} S^z_{{\mb r}+{\mb e}} {\rm e}^{i{\mb r} \cdot {\mb k}},
\end{equation}
where ${\mb e}$ is a unit vector in the $x$ or $y$ direction of the square lattice. Although  $S=2$ states also are excited by $\mathscr{S}_{\mb k}$, a singlet
has the lowest energy among the even-$S$ excited states in this system. Gaps were estimated by the generalized moment method\,\cite{SuwaT2015,SenSS2015}, in which
a series of moments of the imaginary-time  correlation function $\langle \mathscr{T}_{-\mb k}(\tau)\mathscr{T}_{\mb k}(0)\rangle$ or
$\langle \mathscr{S}_{- \mb k}(\tau)\mathscr{S}_{\mb k}(0)\rangle$ is systematically extrapolated to extract the asymptotic exponential decay
time $\tau_{S,{\mb k}}$ (inverse of the gap). The procedures, including error estimation from bootstrap analysis, follow closely our recent work on other systems in the
triplet sector \cite{SenSS2015}. We define $g=J/(J+Q)$ and set $J+Q=1$.

Here we demonstrate our approach for gap estimation. The Fourier transform of the imaginary-time  correlation function is directly 
measured in the QMC simulations,
\begin{align}
\int_0^{\tau_{\rm max} } d\tau \, C(\tau) e^{i \tau \omega_m} = R(\omega_m) + i J(\omega_m) \label{eqn:Fourier-corr},
\end{align}
where $\omega_m = 2 \pi m / \tau_{\rm max} $ $(m \in \mathbf{Z})$, $C(\tau)=\langle \mathscr{T}_{-\mb k}(\tau)\mathscr{T}_{\mb k}(0)\rangle$ 
or $\langle \mathscr{S}_{- \mb k}(\tau)\mathscr{S}_{\mb k}(0)\rangle$, and $R(\omega_m)$ and $J(\omega_m)$ are the real and imaginary parts, 
respectively. The projection length $\beta$ is set long enough to ensure that $C(\tau)$ is properly $\beta \to \infty$ converged for the
relevant values of $\tau$. The series of the gap estimators ($n \geq 1$) is constructed according to
\begin{align}
\hat{\Delta}_{( n,\tau_{\rm max})} = 
- \omega_1^2 \, \frac{ \sum_{m=1}^n x_{n,m,1} \frac{ J(\omega_m) }{ \omega_m } }{ \sum_{m=0}^n x_{n,m,0} R(\omega_m) },
\label{eqn:gap_Fourier}
\end{align}
where $x_{1,1,1}=1$ and otherwise
\begin{align}
x_{n,m,p}= \frac{ 1 }{ \prod_{p \leq j \leq n, j \neq m} ( m + j )( m - j ) }.
\end{align}
The gap ($\Delta$) is then estimated by the extrapolation;
\begin{align}
  \lim_{n, \tau_{\rm max} \rightarrow \infty} \hat{\Delta}_{ (n,\tau_{\rm max}) } = \Delta, \label{eqn:gap_Fourier_limits}
\end{align}
the convergence of which is analytically assured\,\cite{SuwaT2015,SenSS2015}. In practice, the limit $n \to \infty$ is taken first, using results 
for $n \leq 8$, and $\tau_{\rm  max} \to \infty$ is taken subsequently. Here data with relative error bars larger than 1 are excluded (which does not
introduce any bias in the process). The dynamical correlations and the gap extrapolations using 
$\hat{\Delta}(\tau_{\rm max}) \equiv \lim_{n \to \infty} \hat{\Delta}_{(n, \tau_{\rm max})}$ for the $J$-$Q$ model with $J/Q=0.045$, $L=64$, $S=1$, 
are shown in Figs.~\ref{fig:gap-pipi} and~\ref{fig:gap-pi0} for momentum ${\bf k}=(\pi,\pi)$ and $(\pi,0)$, respectively. In the captions of these
figures and in the following, the numbers in parentheses indicate the statistical uncertainty, one standard deviation, on the preceding digit. 
Convergence of the gap estimator is observed in both cases. The statistical precision is high enough to allow the kind of analysis presented in 
the following sections. Note that the data points for different $\tau_{\rm max}$ are correlated, and to take this into account properly, the error 
bars of the final gap estimates are calculated using bootstrapping. 

\begin{figure}
\center{\includegraphics[width=8cm, clip]{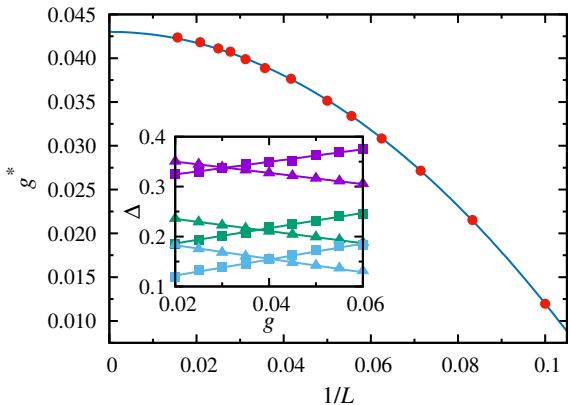}}
\vskip-2mm
\caption{(Color online) Extrapolation of the transition point from the coupling $g^*(L)$ at the crossing between
  the lowest triplet and singlet gaps. A power-law fit, $g^*(L) - g_{\mathrm c} \propto L^{-\sigma}$, for
  $10 \leq L \leq 64$ gives $g_{\mathrm c}=0.04301(8)$ and $\sigma=2.00(1)$ (with $\chi^2/N_{\rm dof} \approx 1.4$).
  The inset shows triplet (triangles) and singlet (squares) gaps for $L$=16, 24, and 32 (top to bottom).}
\label{fig:gap-crossing}
\vskip-2mm
\end{figure}

\section{Level Spectroscopy}
\label{sec:spectroscopy}

We first use level spectroscopy to locate the transition point and extract the critical gap scaling exponent. Different types
of ground states are associated with different low-energy excitations, which can lead to crossings of energy levels with different quantum numbers
as a function of the control parameter used to tune the quantum phase transition. The finite-size scaling of the crossing point provides a remarkably
good estimate of the critical point in several 1D systems\,\cite{Nomura1995,Eggert1996,Sandvik2010c,Sandvik2010b,SuwaT2015}. As for the 2D $J$-$Q$ model,
the triplet excitation is gapless in the N\'eel phase (the lowest triplet being a quantum rotor state with gap scaling as $1/L^2$ \cite{Manousakis1990}),
while it is gapped in the VBS phase. In contrast, the lowest singlet gap of a finite system decreases exponentially with the system size in the VBS phase,
while it converges to a finite value in the N\'eel phase. Therefore, the lowest triplet, which is at $k=(\pi,\pi)$, and singlet, at $k=(\pi,0)$ and 
$(0,\pi)$, cross each other at a coupling which converges to the transition point in the thermodynamic limit.

\begin{figure}
  \center{\includegraphics[width=8cm, clip]{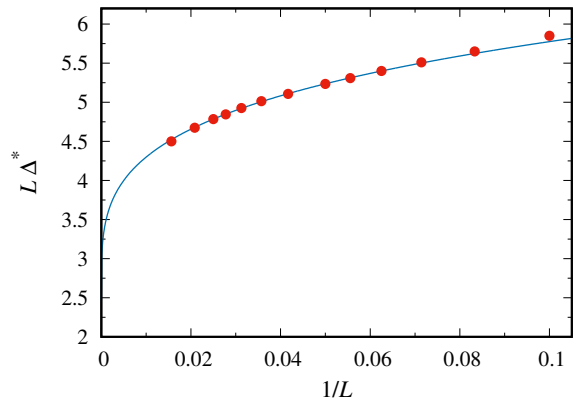}}
\vskip-2mm
\caption{(Color online) Scaling of the gap at the crossing point. A fit to the form $L\Delta^* = a + b L^{-\tau}$ for
  $16 \leq L \leq 64$ gives $\tau=0.26(4)$ (with $\chi^2/N_{\rm dof} \approx 1.8)$.}
\label{fig:scaling}
\vskip-2mm
\end{figure}

Figure~\ref{fig:gap-crossing} presents our results, with examples of level crossings shown in the inset and the finite-size drift of the 
crossing points $g^*(L)$ analyzed in the main figure. The crossing points have been fitted to a constant (the infinite-size critical point) with
a power-law correction $\propto L^{-\sigma}$, with $\sigma=2.00(1)$. The critical point $g_{\mathrm c}=g^*(L\to \infty)=0.04301(8)$, or 
$(J/Q)_{\rm c}=0.04494(9)$, is in reasonably good agreement with a recent, more precise estimate $(J/Q)_{\rm c}=0.04468(4)$\,\cite{ShaoGS2016}. 
Before discussing the information contained in the correction exponent $\sigma$, in Fig.~\ref{fig:scaling} we present data for the gap at the crossing 
point. Given that the expected dynamic exponent $z=1$, we here graph the crossing gap $\Delta=\Delta_s=\Delta_t$ multiplied by the system size $L$, 
and again fit with a power-law correction; $\propto L^{-\tau}$ with $\tau=0.26(4)$.

Given the above results and the scaling hypothesis introduced in \cite{ShaoGS2016}, we analyze the scaling of the lowest triplet ($\mu=t$)
and singlet ($\mu=s$) gaps $\Delta_{\mu} (\delta,L)$ with the distance $\delta=g-g_\mathrm{c}$ from the DQC point using
\begin{equation}
  \Delta_{\mu} (\delta,L) = L^{-1} f_\mu (\delta L^{1/\nu}, \delta L^{1/\nu'}, L^{-\omega} ),
   \label{eqn:scaling}
\end{equation}
where $\nu \approx 0.45$ and $\nu'\approx 0.58$ are the values from Ref.~\cite{ShaoGS2016} of the exponents
governing the correlation length and the U($1$) scale, respectively, and $\omega$ is the exponent of the leading irrelevant field
(for which a small value, $\omega \approx 0.3$, was found in scaling of other quantities in Ref.~\cite{ShaoGS2016}). The functions $f_\mu$ should 
approach constants when $\delta \to 0$, up to additive size corrections from the $L^{-\omega}$ dependence (and higher-order corrections not 
included here). 

To analyze the finite-size scaling of the gaps, we begin in the standard way by Taylor expanding the postulated scaling functions $f_\mu$
in Eq.~(\ref{eqn:scaling}) to leading order in the relevant and irrelevant fields. For the singlet and triplet cases we have 
\begin{eqnarray}
  L\Delta_s & = & a_s + b_s\delta L^{1/\nu}+ c_s\delta L^{1/\nu'}+d_sL^{-\omega_s},\label{tay1} \\
  L\Delta_t & = & a_t + b_t\delta L^{1/\nu}+ c_t\delta L^{1/\nu'}+d_tL^{-\omega_t}, \label{tay2}
\end{eqnarray}
where we have used the expected value of the dynamic exponent, $z=1$, and allow for the possibility of different correction exponents, $\omega_s$ 
and $\omega_t$, for the two gaps. In principle the leading irrelevant corrections could arise from the ratio $L^{1/\nu'-1/\nu}$ of the arguments 
$\delta L^{1/\nu}$ and $\delta L^{1/\nu'}$ of $f_\mu$, in which case we can just replace the exponents $\omega_s$ or $\omega_t$ as appropriate 
by $1/\nu-1/\nu'$.

We are interested in the crossing point of the scaled gaps, the value of $\delta=g-g_c$ for which $L\Delta_s=L\Delta_t$.
Defining $a=a_s-a_t$ , $b=b_t-b_s$, and $c=c_t-c_s$, we obtain the crossing point $\delta^*_s(L)=g^*(L)-g_c$ as a function
of the system size:
\begin{equation}
\delta^*(L) = \frac{a+d_sL^{-\omega_s} - d_tL^{-\omega_t}}{bL^{1/\nu}+cL^{1/\nu'}} \label{delta}.
\end{equation}  
In general, if $a\not=0$, i.e., if the scaled gaps are different at the critical point when $L \to \infty$, we see that the crossing
point for large $L$ drifts as $\delta^*(L) \propto L^{-1/\nu}$, provided also that the coefficient $b\not=0$. This is not consistent
with the observation that $\delta^*(L) \propto L^{-\sigma}$ with $\sigma \approx 2.00$ (Fig.~\ref{fig:gap-crossing}),
given that $1/\nu \approx 2.25$. If $b=0$, we have $\delta^*(L) \propto L^{-1/\nu'}$, but this is also not consistent with the data,
because $1/\nu' \approx 1.71$. However, if $a=0$ and $b=0$, we have $\delta^*(L) \propto L^{-1/\nu'-\omega}$, where $\omega$ is
the smaller of $\omega_s$ and $\omega_t$. Then $\sigma=1/\nu'+\omega=2.00(1)$ and $\tau=\omega=0.26(4)$. This case is fully compatible with the data; $1/\nu' = \sigma - \tau = 1.74(4)$, in excellent agreement with the previous value $1/\nu'=1.71(3)$ \cite{ShaoGS2016}. The value of $\omega$ is also in good agreement with a renormalization-group calculation within the field theory~\cite{Bartosch2013}. 

\begin{figure}
\center{\includegraphics[width=8cm, clip]{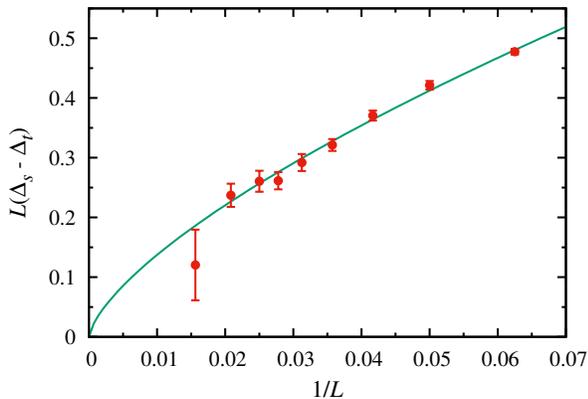}}
\vskip-2mm
\caption{(Color online) Power-law scaling of the difference between the singlet and triplet gaps. The curve is a fit to $ \propto L^{-\omega'}$,
where $\omega' =0.68$.}
\label{fig-lowest-gaps}
\vskip-2mm
\end{figure}

We here also present more direct evidence for $a_s=a_t$ in Eqs.~(\ref{tay1}) and (\ref{tay2}).
Figure~\ref{fig-lowest-gaps} shows that the difference $L(\Delta_s-\Delta_t)$ between the scaled
gaps for $J/Q=0.045$ very close to the estimated $g_c$ goes to zero, along with a power-law fit giving a correction exponent $\omega'=0.68(3)$. It is
interesting that this exponent is approximately twice the value of the leading correction exponent $\omega \approx 0.3$ that we have found for
other quantities. It is then plausible that $\omega'$ corresponds to the quadratic contributions from the leading irrelevant
field, i.e., $\omega'=2\omega$. Another possibility is that $\omega'=1/\nu-1/\nu'$, which is also consistent with the known values of the exponents
$\nu$ and $\nu'$, which give $1/\nu-1/\nu' \approx 0.6$. Within the current statistical precision, $2\omega$ and $1/\nu-1/\nu'$ cannot be distinguished.

The higher-power-law scaling ($\omega' > \omega$) could apparently mean that the leading corrections of the two gaps are equal, $\omega_s=\omega_t$, and the prefactors of the power laws are the same, that is, $d_s L ^{-\omega_s} - d_t L^{ - \omega_t} = 0$. This perfect cancellation, however, leads to $\sigma = 1/\nu' + \omega'$, which is not consistent with data; thus the leading terms should not be canceled perfectly. What is surprising here is rather that the dominant contribution ($\omega'$ term) to the difference of the scaled gaps seems absent in the crossing-coupling ($\delta^*$) scaling, as shown in Fig.~\ref{fig:gap-crossing}. This nontrivial cancellation implies a relation in the correction terms although it is not easy to identify among many possibilities.

\begin{figure}
\center{\includegraphics[width=8cm, clip]{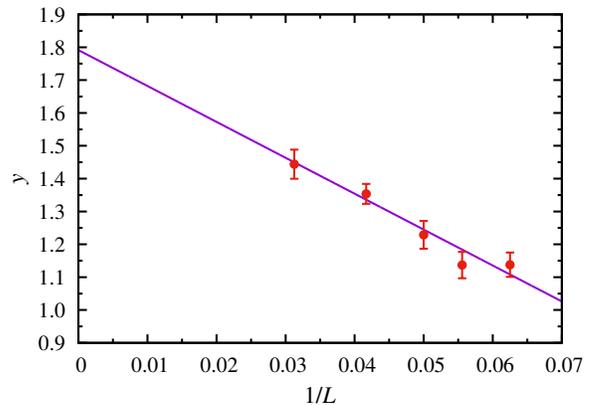}}
\vskip-2mm
\caption{(Color online) Estimates of the inverse of the exponent controlling the triplet gap away from the critical point, extrapolated to
infinite size using a linear fit.}
\label{fig-nup}
\vskip-2mm
\end{figure}

An estimate for the exponent controlling the scaling of the gap away from the critical point can be obtained by using the derivative
of the scaling functions $f_\mu$ with respect to $\delta$. Let us define the scaled derivative of the triplet gap
\begin{equation}
D_t(L)= L \frac{d\Delta_t}{d \delta}= L \frac{d\Delta_t}{d g}.
\end{equation}
From the Taylor expansion in Eq.~(\ref{tay2}) we obtain the leading-order behavior
\begin{equation}
\ln [D_t(L)] = \ln(b_\mu) + \frac{1}{\tilde \nu}\ln(L),
\end{equation}
where $\tilde \nu$ is the exponent ($\nu$ or $\nu'$) controlling the scaling.
We can devise an estimator for the exponent controlling the relevant length scale:
\begin{equation}
y(L) = \frac{\ln[D_t(2L)] - \ln[D_t(L)]}{\ln(2)}.
\end{equation}
This estimator converges to $1/\nu$ if $b_t \neq 0$, or to $1/\nu'$ if $b_t=0$ and $c_t \neq 0$, in Eq.~(\ref{tay2}), when $L \to \infty$, since $1/\nu > 1/\nu'$.
Figure~\ref{fig-nup} shows $y(L)$ graphed versus $1/L$. Here we estimated the derivative by linear interpolation of the gaps
around the crossing point between the lowest singlet and triplet excitations, as shown in Fig.~\ref{fig:gap-crossing}.
We expect that the lowest triplet gaps in the VBS phase should be governed by the emergent U$(1)$ length scale, i.e., the exponent $\nu'$,
based on the finding in Ref.~\cite{ShaoGS2016} that the length scale of triplets (the confinement length scale) diverges with this
exponent. Then the general form (\ref{tay2}) of the gap must have $b_t=0$ and $c_t \neq 0$. We do not have enough data in Fig.~\ref{fig-nup} to
meaningfully analyze (with error bars sufficiently small for the results to be useful)
the behavior with a power-law form with an adjustable power. The behavior appears to be essentially linear in $1/L$, however, which
is similar to the scaling of the triplet length scale observed in Ref.~\cite{ShaoGS2016}. For a rough estimate, we therefore simply perform
a straight-line fit. This gives $y(\infty)=1.79(7)$, which is consistent with the previous value of $1/\nu'$.

\begin{figure}
\center{\includegraphics[width=8cm, clip]{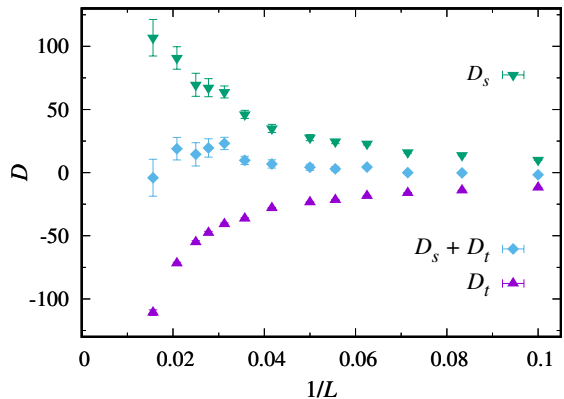}}
\vskip-2mm
\caption{(Color online) Scaling of the derivatives of
  the singlet and the triplet gaps with respect to $g$, along with the slower-divergent sum of the two.}
\label{fig:derivative}
\vskip-2mm
\end{figure}

In the thermodynamic limit, the scaling form (\ref{eqn:scaling}) when $L \to \infty$ is compatible with $\Delta \propto \delta^{z\nu}$ 
if $f_\mu \to (\delta L^{1/\nu})^{z\nu}$ or $ \Delta \propto \delta^{z\nu'}$ if $f_\mu \to (\delta L^{1/\nu'})^{z\nu'}$. As we have inferred, from the gap-crossing scalings, that the
finite-size scaling (where the first two arguments of $f_\mu$ are small) is governed by $y=\delta L^{1/\nu'}$ although $x=\delta L^{1/\nu}$ is the larger
argument, $x$ should not even appear in the scaling function (except possibly in a ratio $y/x$, which acts as an irrelevant field). Thus, we suspect
that both the singlet and triplet gaps scale as $\delta^{z\nu'}$ near the critical point.
This is also physically plausible because there should be states (singlets and triplets)
above the four degenerate VBS singlets for $g<g_c$ related to the emergent $U(1)$ gauge field, which in the DQC scenario is governed
by $\nu'$ \cite{SenthilVBSF2004,ShaoGS2016}. In the N\'eel phase as well there should be a singlet energy reflecting the longer length scale \cite{SenthilVBSF2004}.

Note that for quantities whose finite-size scaling is governed by the argument $x=\delta L^{1/\nu}$ in Eq.~(\ref{eqn:scaling}), the $L\to \infty$ form of
$f_\mu$ can still be governed by $y=\delta L^{1/\nu'}$, and this behavior can be associated with anomalous finite-size powers \cite{ShaoGS2016}.
In the case of the gaps, we here instead found conventional finite-size scaling, $\Delta \propto L^{-z}$, but a thermodynamic limit controlled
by $\nu'$.

As shown in the inset of Fig.~\ref{fig:gap-crossing}, the gap is linear in $L$ for $g$ close to $g_{\mathrm c}$.
The gap derivatives $D_{\mu}(L) \equiv L \, \partial \Delta_{\mu}(g,L) / \partial g$ are shown in Fig.~\ref{fig:derivative}, exhibiting divergences with opposite signs.
We also show the scaling of $D_s + D_t$, where it appears that the leading divergence is canceled, with
only a weaker divergence remaining (which should be $\propto L^{1/\nu'-\omega}$). The cancellation, together with the identification of the scaled gap values observed in Fig.~\ref{fig-lowest-gaps}, implies that the scaling functions for $L \to \infty$ are
symmetric, at least in the linear order, around the critical point: $f_s(y)=f_t(-y)$ with $y=\delta L^{1/\nu'}$. This duality between singlet and triplet excitations supports the
proposed SO($5$) symmetry at the critical point\,\cite{NahumSCOS2015}. Note that with the three triplets at $q=(\pi,\pi)$ and singlets at $(\pi,0)$
and $(0,\pi)$, we have a total of five gapless modes that scale in the same way.

\section{Linearly Dispersing Spinons}
\label{sec:lineardisp}

We next show the dispersion curve around the DQC point and investigate carefully the finite-size scaling of the lowest gaps. It will be shown that at the critical point, there are both gapless singlets and triplets at ${\bf k}=(0,0), (\pi,0), (0,\pi)$, and $(\pi,\pi)$. In addition, the velocities around the gapless modes are studied by the winding-number method and the direct-gap measurement. It is found that a unique velocity appears around the multiple gapless modes. These findings strongly indicate that the low-energy excitation is formed by linearly dispersing spinons.

\subsection{Full Dispersion}
\begin{figure}[t]
\center{\includegraphics[width=8cm, clip]{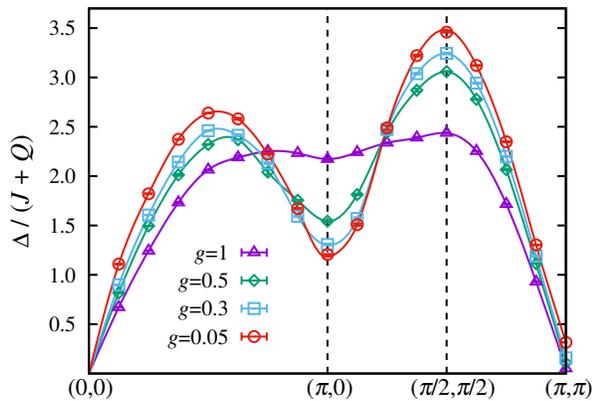}}
\vskip-2mm
\caption{(Color online) Dispersion relation of the lowest triplet for $L=16$ systems at several coupling parameters. The
splines connecting points are only guides to the eye.}
\label{fig:dispersion}
\vskip-2mm
\end{figure}

We study the dispersion of the lowest triplet. Figure~\ref{fig:dispersion} shows results for $L=16$ and several values of $g$ along
a standard path in the Brillouin zone. Spin-wave theory for the Heisenberg model produces a magnon-excitation energy that is in generally good
agreement with numerical calculations \cite{Manousakis1990}. The main discrepancy is at the AFM zone boundary, the line from $k=(\pi/2,\pi/2)$ to
$(\pi,0)$, where in spin wave theory to order $1/S$ there is no dispersion. Numerical calculations show a $10\%$ lower energy at $(\pi,0)$
\cite{Singh1995,Sandvik2001}. The minimum at $(\pi,0)$ has been termed the ``roton minimum'' \cite{Powalski2015}, in analogy with the local dispersion
minimum in $^4$He; it was argued that it originates from interactions between the transverse (magnon) and longitudinal (``Higgs'') modes.
In the $J$-$Q$ model, we can see that the differences between $k=(\pi/2,\pi/2)$ and $(\pi,0)$ increase dramatically as we approach the critical point.
The reduction in $(\pi,0)$ energy is in accord with a variational argument\,\cite{SpanuBS2006}, according to which the triplet gap at $(\pi,0)$ must vanish
if the N\'eel-to-VBS transition is continuous. For the relatively small system in Fig.~\ref{fig:dispersion}, the expected gapless $(\pi,0)$ mode 
is not yet apparent and requires a finite-size analysis close to $g_c$, as we discuss below. Even with the data in Fig.~\ref{fig:dispersion}, it is now
clear that the weak roton minimum of the Heisenberg model is due to VBS fluctuations (which should also be related to emergent gauge
bosons \cite{Huh2013}), as also discussed in \cite{Ghioldi2016}, which are strengthened as $Q/J$ is increased and 
push the minimum down to $0$ as the DQC point is approached.

\subsection{Finite-size Scaling}
\begin{figure}[t]
\center{\includegraphics[width=8.2cm, clip]{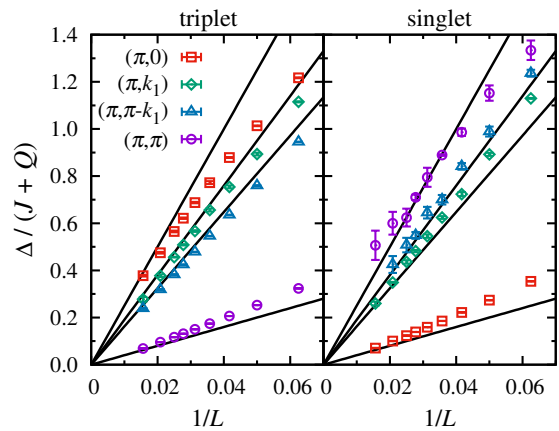}}
\vskip-2mm
\caption{(Color online) Triplet (left) and singlet (right) gaps at several wave vectors for systems close to the DQC point ($J/Q=0.045$). 
The smallest wave-vector increment $2\pi/L$ is denoted by $k_1$. The solid lines illustrate the expected critical form $\Delta \propto L^{-1}$
with $\Delta_{t,(\pi,q)}= \Delta_{s,(\pi, \pi - q)}$ imposed.} 
\label{fig:gaps}
\vskip-2mm
\end{figure}

In addition to the gapless triplet at ${\bf k}=(\pi,\pi)$ and singlets at $(\pi,0)$, $(0,\pi)$, we also expect a gapless triplet at ${\bf k}=(0,0)$, as is well known
in the Heisenberg model and which is also reflected in Fig.~\ref{fig:dispersion}. Next we will show that actually there are both gapless singlets and
triplets at all these four points, ${\bf k}=(0,0), (\pi,0), (0,\pi)$, and $(\pi,\pi)$; that is, there are {\it eight} gapless excitation modes in total. We focused the calculations at $J/Q=0.045$, close to the estimated
critical value, and extracted the gaps $\Delta_{\mu,\mb k}$ at several wave vectors. Results are displayed in Fig.~\ref{fig:gaps}, along with
fits to the expected $1/L$ forms for $z=1$ criticality. A very interesting observation is a singlet-triplet symmetry---a generalization of the equivalence of
the lowest $s$,$t$ gaps: $\Delta_{t,(\pi,q)} \approx \Delta_{s,(\pi, \pi - q)}$ is seen for $q=0,2\pi/L, \pi - 2\pi/L$, and $\pi$. The singlets are a bit higher than the
corresponding triplets, likely because of higher-order irrelevant fields as the differences appear to vanish as $L \to \infty$.
In Fig.~\ref{fig:gaps} we draw lines with the same prefactors in $1/L$ for the corresponding
gaps. These findings strongly suggest that the system has gapless singlet and triplet excitations at $(0,0)$, $(\pi,0)$, $(0,\pi)$, and $(\pi,\pi)$,
with a remarkable relationship between the finite-size corrections for singlets and triplets that may again be related to emergent SO(5) symmetry.

\begin{figure}
\center{\includegraphics[width=8cm, clip]{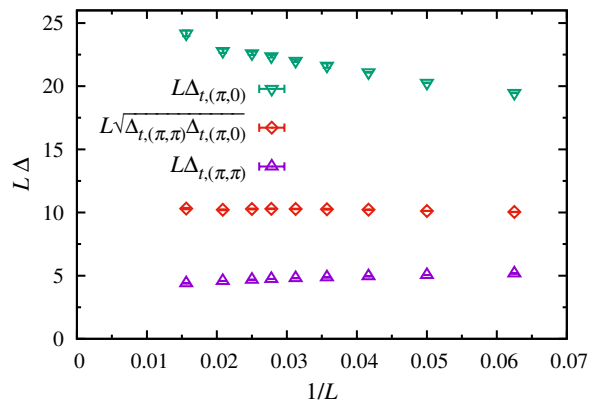}}
\vskip-2mm
\caption{(Color online) Scaling of the triplet gaps at ${\bf k}=(\pi,\pi)$ and $(\pi,0)$ and of the square root of the product of the two.
 The coupling ratio is $J/Q=0.045$, close to the estimated critical coupling.}
\label{fig-sqprod}
\vskip-2mm
\end{figure}

Another cancellation of corrections is found in an analysis of the triplets at $(\pi,\pi)$ and $(\pi,0)$. Here an almost
perfect cancellation of corrections is seen in the product of the gaps: $\Delta_{t,(\pi,\pi)}\Delta_{t,(\pi,0)}$. In Fig.~\ref{fig-sqprod}
the two gaps along with the square root of the product are scaled by $L$. The individual scaled gaps approach finite values as $L \to \infty$,
but there are significant corrections. The corrections in the product are much smaller and not seen on the scale of Fig.~\ref{fig-sqprod}, indicating
that the $(\pi,\pi)$ and $(\pi,0)$ gaps scale as $L^{-1}(1 \pm aL^{-\omega})$ with different signs in front of the $L^{-\omega}$ correction.
Then the leading corrections
cancel in the product and the remaining correction is $\propto a^2 L^{-2\omega}$, which, apparently, is overall too small to clearly see in Fig.~\ref{fig-sqprod}
(likely because of a very small prefactor $a^2$).

It should be noted that the coupling ratio considered here, $J/Q=0.045$, is very close to but not exactly at the critical ratio
$(J/Q)_c \approx 0.0447$ \cite{ShaoGS2016}. Being slightly on the AFM side of the transition, the triplet $(\pi,\pi)$ gap could be marginally
affected by the quantum rotor states, the gaps to which asymptotically scale as $L^{-2}$ \cite{Manousakis1990}, and the $(\pi,0)$ triplet may be
affected by the small gap expected in the weak AFM state at this wave vector. The corrections analyzed above could then be partially due to crossovers into such AFM
scaling, i.e., the exponent $\omega$ would then be an effective exponent only for the range of sizes considered. Analyzing the gaps
at the crossing point, as we did in Fig.~\ref{fig:scaling}, avoids this issue, and since the value of $\omega$ obtained there is
very similar to what can be inferred from below in Fig.~\ref{velocity-winding-number} for a different
quantity computed at $J/Q=0.045$, we conclude that the effects from not being exactly at the critical point should be very minor here.

\subsection{Unique Velocity}

We present an estimate of the velocity obtained using winding numbers first, and thereafter discuss the more direct approach using gaps.
For a system with conserved magnetization, the standard QMC mappings from the partition function in $d$ dimensions to an effective one with
$d+1$ dimensions leads to topologically conserved winding numbers. The temporal winding number $W_\tau$ simply counts the difference
between the number of up and down spins, while the spatial winding numbers $W_r$ (here $r=x,y$ in two dimensions) correspond to a quantization of
the net spin currents due to periodic boundaries in space and time. If a QMC simulation includes updates that can change the winding numbers,
which the loop updates in the stochastic series expansion (SSE) method used here indeed can
\cite{Sandvik2010b}, physical quantities related to the winding-number fluctuations
can be computed \cite{Pollock1987}. The uniform
magnetic susceptibility is given by
\begin{equation}
\chi= \frac{\beta}{N} \left \langle M_z^2 \right \rangle = \frac{\beta}{N}\left \langle W_\tau^2 \right \rangle,
\label{ususc}
\end{equation}
where $M_z$ is the total magnetization. In two dimensions the spin stiffness is given by
\begin{equation}
\rho_s = \frac{1}{2\beta}\left ( \left \langle W_x^2 \right \rangle  + \left \langle W_y^2 \right \rangle  \right ).
\label{rhos}
\end{equation}
For a many-body system with dynamic exponent $z=1$ (linear dispersion), Lorentz invariance is emergent when $L \to \infty$ and $\beta =1/T \to \infty$. The effective
length of the system in the time dimension is $L_\tau=c\beta$. It has been argued that a cubic space-time geometry (i.e., with the system having effectively
equal lengths in space and time) should be defined by requiring the following condition for a given spatial system size $L$ \cite{Jiang2011}:
\begin{equation}
\left\langle W_r^2(\beta^*) \right\rangle = \left \langle W_\tau^2 (\beta^*) \right\rangle,
\label{cubic}
\end{equation}
where $\beta^*$ is the unique value of $\beta$ for which the equality holds. This criterion offers an interesting way to compute the velocity
$c$ of excitations as the aspect ratio $c=L/\beta^*$. One can expect this procedure to deliver the correct velocity in the limit $L \to \infty$.
In some cases this can be shown directly based on low-energy field theory \cite{Kaul2008,Jiang2011}, but even
in the absence of such descriptions the arguments are very general and one can expect the correct velocity for any system with linear dispersion.
In Ref.~\cite{SenSS2015} we presented several high-precision tests for both AFM ordered and critical systems.

\begin{figure}
\center{\includegraphics[width=7.4cm, clip]{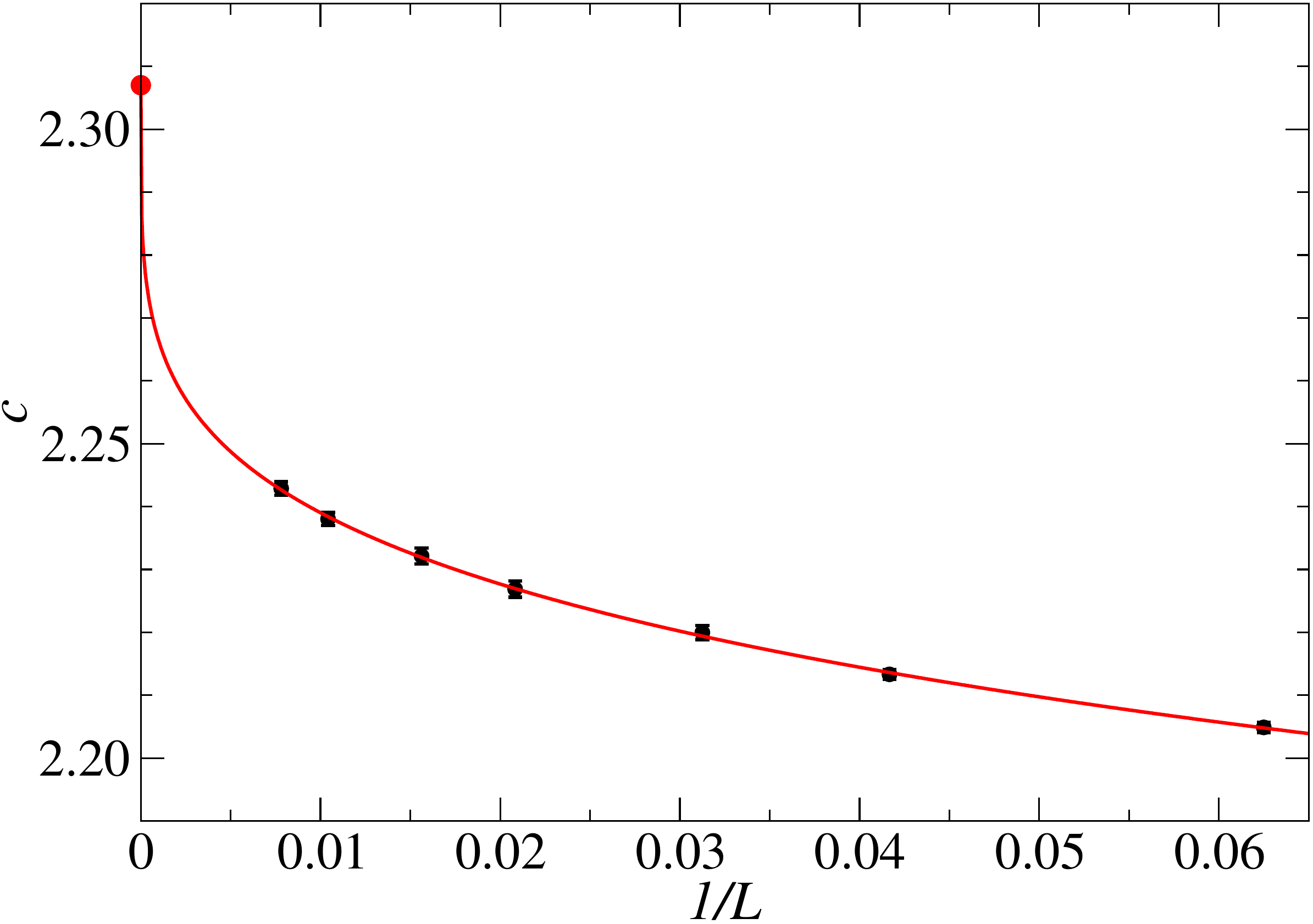}}
\vskip-2mm
\caption{(Color online) Velocity estimates for the $J$-$Q$ model at $J/Q=0.045$ [close to its critical point, $(J/Q)_c \approx 0.447$] extracted using the cubic criterion
  Eq.~(\ref{cubic}). The unit of $c$ corresponds to setting the lattice constant to $1$ and the energy scale $J+Q=1$. The curve shows a fit including
  a power-law correction $\propto L^{-\omega}$ to the infinite-size velocity. The fit with error analysis gives $c=2.31(5)$ and $\omega=0.24(8)$.}
\label{velocity-winding-number}
\vskip-2mm
\end{figure}

\begin{figure}
\center{\includegraphics[width=8cm, clip]{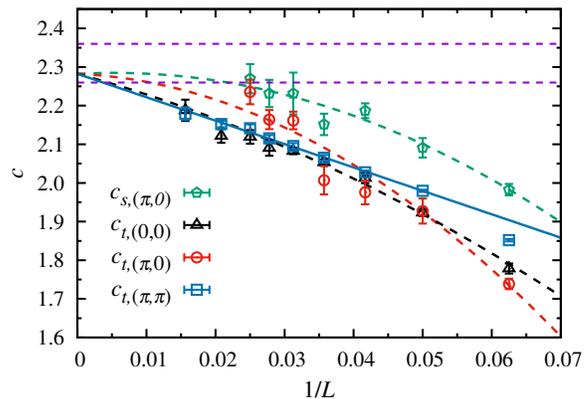}}
\vskip-2mm
\caption{(Color online) Size-dependent velocity estimators around the gapless points. The solid (blue) line is a fit to the $c_{t,(\pi,\pi)}$ data
  for $L \geq 20$ and gives $c=2.282(5)$ ($\chi^2/N_{\rm dof} \approx 0.4$). The fits to the other data sets use the same $c$
  along with $L^{-1}$ and $L^{-2}$ corrections. The horizontal dashed lines indicate the value of $c$ $\pm$ one standard deviation obtained using
  a winding-number estimator.}
\label{fig:velocity}
\vskip-2mm
\end{figure}

We have used the winding-number method also for the $J$-$Q$ model and present results for several system sizes at $J/Q=0.045$ in Fig.~\ref{velocity-winding-number}.
The finite-size data are in excellent agreement with a constant plus a finite-size correction $\propto L^{-\omega}$, with the velocity $c/(J+Q)=2.31(5)$
in the thermodynamic limit and $\omega=0.24(8)$. As shown in Fig.~\ref{fig:velocity}, the value of $c$ is in excellent agreement with
the velocity extracted using energy gaps, and it is also in good agreement with a previous QMC calculation (where, however, no scaling correction was used in the
analysis) \cite{Kaul2008}. We also note that the value of the correction exponent $\omega$ is close to values extracted based on other
quantities, here as well as in Ref.~\cite{ShaoGS2016}, and it is also close to a result based on a renormalization-group calculation within the
the DQC field theory \cite{Bartosch2013}.

We next extract the velocity $c$ of excitations, using ${\bf k}$ points away from the gapless points ${\bf K}_0$ by amounts $k_1,k_2$,
where  $k_n=n2\pi/L$:
\begin{equation}
  c_{\mu,\mb{K}_0}(L) \equiv ({L}/{2\pi}) \left[ \Delta_{\mu,\mb{K}_0+\mb{k}_2}(L) - \Delta_{\mu, \mb{K}_0+\mb{k}_1}(L) \right].
  \label{eqn:c}
\end{equation}
At $(\pi,0)$ we have two options for the direction of the small displacements $k_n$, and we find the best statistical precision with
${\mb K}_0+{\mb k}_n=(\pi,k_n)$ and $(\pi-k_n,0)$ for the triplet and singlet, respectively. In principle we can also define the velocity based solely
on the gaps $\Delta_{\mu,\mb{K}_0+\mb{k}_1}$, but Eq.~(\ref{eqn:c}) has smaller size corrections. For ${\mb k}\not=0$, we expect momentum-dependent
corrections in the form of integer powers of $1/L$ \,\cite{SenSS2015},
\begin{equation}
  \Delta_{\mu,\mb{K}_0+\mb{k}}(L) = c k + B_{\mu,\mb{K}_0+\mb{k}}L^{-1} + O(L^{-2}),
  \label{eqn:gap}
\end{equation}
i.e., the nontrivial critical scaling behavior is seen only exactly at the gapless points (${\mb k}=0$),
and the coefficients can be expanded as $B_{\mu,\mb{K}_0+\mb{k}} = a_{\mu,\mb{K}_0}+b_{\mu,\mb{K}_0}\,k + O(k^2)$.
Note that $\lim_{\mb{k} \to 0}B_{\mu,\mb{K}_0+k} \neq B_{\mu,\mb{K}_0}$ in general \cite{SenSS2015}.

The estimator $c_{\mu,\mb{K}_0}(L) = c + b_{\mu,\mb{K}_0}/L+O(1/L^2)$ converges to the correct velocity in the thermodynamic limit even at a
critical point as long as $z=1$ \cite{SenSS2015}. We here analyze those singlets and triplets for which the gaps were determined to sufficient
precision. As shown in Fig.~\ref{fig:velocity}, the velocities appear to converge to the same value. We have the highest precision for the triplet
at $\mb{K}_0=(\pi,\pi)$, giving $c=2.282(5)$. For the other cases we simply fit curves with this $c$ fixed.
The velocity estimates from the winding-number method and the direct gap measurements agree within statistical error. The velocities of the 
linearly dispersing modes for both singlets and triplets around all four gapless points being equal to each other again points to an emergent 
symmetry between the low-lying singlets and triplets at the critical point.

\begin{figure}[t]
\center{
\includegraphics[scale=0.35, bb=0 0 620 180, clip]{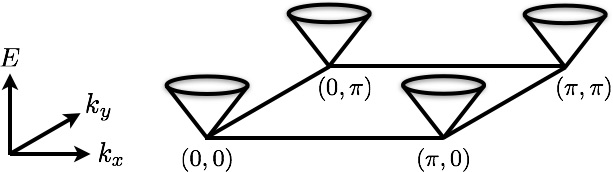}}
\vskip-2mm
\caption{Schematic illustration of the low-lying energy spectrum at the deconfined quantum critical point. There are four gapless points 
for both $S=0$ and $S=1$ excitations, at $\mb{k}=(0,0), (\pi,0), (0,\pi),$ and $(\pi,\pi)$. Close to these points the modes disperse linearly 
with the same velocity. The dispersion relation marks the lower edge of a continuum of excitations arising from two essentially deconfined
spinons, with the single-spinon dispersion also being the same as the lower edge of the two-spinon continuum.}
\label{fig:schematic}
\vskip-2mm
\end{figure}

\section{Summary and Conclusions}
\label{sec:summary}

We have studied the excitation gaps and the dispersion relation of the SU(2) symmetric $J$-$Q$ model on the square lattice using unbiased QMC methods. 
The transition point was located by level spectroscopy of the lowest excitation gaps (locating gap crossing points), which correspond to triplet 
excitations at ${\bf k}=(\pi,\pi)$ and singlets at ${\bf k}=(\pi,0)$ and $(0,\pi)$. We found a duality of the scaling function governing these gaps
and estimated the relevant critical exponent governing the gap scaling in the ordered phases, which we argued is the same exponent, $\nu'$, that 
governs the emergent U(1) symmetry in the VBS phase, i.e., not the standard correlation length exponent $\nu$. The value of $\nu'$ is consistent 
with other recent estimates \cite{ShaoGS2016}, as is the leading irrelevant exponent, $\omega \approx 0.3$. 

At the critical point, there are both gapless singlets and triplets at ${\bf k}=(0,0), (\pi,0), (0,\pi)$, and $(\pi,\pi)$, forming {\it eight} gapless 
excitation modes, in total, with the same velocity, as illustrated in Fig.~\ref{fig:schematic}. The unique velocity for these gapless modes and the 
degenerate singlets and triplets clearly point to deconfined spinon excitations. From the fact that we have measured two-spinon excitation from the 
ground state with $S=0$ and ${\bf k}=(0,0)$ via the projector algorithm which uniquely singles out the ground state with momentum ${\bf k}=(0,0)$, we infer 
that the single-spinon dispersion relation is equal to the degenerate $S=0$ and $S=1$ dispersions at criticality (i.e., close to the $g=0.05$ curve in 
Fig.~\ref{fig:dispersion}, which is already almost $L \to \infty$ converged away from the gapless points). This situation is a direct analog
of the excitations of the $S=1/2$ Heisenberg chain.

In the DQC theory, weak spinon-spinon 
interactions are predicted to lead to different scaling prefactors between the corresponding singlets and triplet (or possibly a finite gap for 
singlets, which we do not find any indication of here)~\cite{SenthilBSVF2004}. To within our numerical precision, the dominant $1/L$ prefactor of 
the gap scalings are the same, although, of course, we cannot rule out very small differences. 
Our study therefore suggests that the spinons in fact are fully deconfined in the case of SU(2) spins, and that this may be 
directly related to an emergent SO(5) symmetry \cite{Nahum2015b}, which has been argued to be a special DQC feature not present for SU(N) spins with $N>2$.

\acknowledgments
{H.S. thanks Cristian Batista for discussion on spinon deconfinement. He acknowledges support by KAKENHI under Grant No.~16K17762, 
a Postdoctoral Fellowship for Research Abroad from JSPS, and a Sasakawa Scientific Research Grant from The Japan Science Society. He also 
acknowledges the computational resources of the Supercomputer Center at the Institute for Solid State Physics, the University of Tokyo, and would like to thank Boston University's Condensed Matter Visitors Program for their support. A.S. 
acknowledges discussions with R. Moessner and the hospitality of the Max Planck Institute for the Physics of Complex Systems (MPIPKS),
Dresden, during the final stages of this work, through support of the Partner Group program between the Indian Association for the Cultivation of Science, Kolkata and MPIPKS and the Visitors program of MPIPKS. A.W.S. was supported by the NSF under Grant No.~DMR-1410126. Some of the calculations were carried out on Boston University's Shared Computing Cluster.}


\begin{thebibliography}{99}

\bibitem{Fisher1989}
  M. P. A. Fisher, P. B. Weichman, G. Grinstein, and D. S. Fisher, 
  Phys. Rev. B {\bf 40}, 546 (1989).
  
\bibitem{ChubukovSY1994}
  A. V. Chubukov, S. Sachdev, and J. Ye, Phys. Rev. B {\bf 49}, 11919 (1994).

\bibitem{Sandvik2002}
  A. W. Sandvik, S. Daul, R. R. P. Singh, and D. J. Scalapino,
  Phys. Rev. Lett. {\bf 89}, 247201 (2002).

\bibitem{Motrunich2004}
  O. I. Motrunich and A. Vishwanath,
  Phys. Rev. B {\bf 70}, 075104 (2004).

\bibitem{SenthilVBSF2004}
  T. Senthil, A. Vishwanath, L. Balents, S. Sachdev, and M. P. A. Fisher,
  Science {\bf 303}, 1490 (2004).  

\bibitem{SenthilBSVF2004}  
  T. Senthil, L. Balents, S. Sachdev, A. Vishwanath, and M. P. A. Fisher,
  Phys. Rev. B {\bf 70}, 144407 (2004).

\bibitem{ReadS1990}
  N. Read and S. Sachdev, Phys. Rev. Lett. {\bf 62}, 1694 (1989);
  Phys. Rev. B {\bf 42}, 4568 (1990).

\bibitem{Sandvik2007}
  A. W. Sandvik,
  Phys. Rev. Lett. {\bf 98}, 227202 (2007).

\bibitem{Manousakis1990}
  E. Manousakis, Rev. Mod. Phys. {\bf 63}, 1 (1991).
  
\bibitem{Wang2011}
  L. Wang, Z.-C. Gu, F. Verstraete, and X.-G. Wen, arXiv:1112.3331. 
  
\bibitem{LiBHS2012}
  T. Li, F. Becca, W. Hu, and S. Sorella,
  Phys. Rev. B {\bf 86}, 075111 (2012).

\bibitem{GongZSMF2014}
  S.-S. Gong, W. Zhu, D. N. Sheng, O. I. Motrunich, and M. P. A. Fisher,
  Phys. Rev. Lett. {\bf 113}, 027201 (2014).

\bibitem{Sandvik2010b}
  A. W. Sandvik, AIP Conf. Proc. {\bf 1297}, 135 (2010).

\bibitem{MelkoK2008}
   R. G. Melko and R. K. Kaul, Phys. Rev. Lett. {\bf 100}, 017203 (2008).

 \bibitem{JiangNCW2008}
   F. J. Jiang, M. Nyfeler, S. Chandrasekharan, and U. J. Wiese,
   J. Stat. Mech. (2008) P02009.

 \bibitem{Sandvik2010a}
   A. W. Sandvik,
   Phys. Rev. Lett. {\bf 104}, 177201 (2010).
   
\bibitem{Kaul2011}
   R. K. Kaul, Phys. Rev. B {\bf 84}, 054407 (2011).

 \bibitem{HaradaSOMLWTK2013}
  K. Harada, T. Suzuki, T. Okubo, H. Matsuo, J. Lou, H. Watanabe, S. Todo, and N. Kawashima,
  Phys. Rev. B {\bf 88}, 220408(R) (2013).

\bibitem{ChenHDKPS2013}
  K. Chen, Y. Huang, Y. Deng, A. B. Kuklov, N. V. Prokof’ev, and B. V. Svistunov,
  Phys. Rev. Lett. {\bf 110}, 185701 (2013).

\bibitem{BlockMK2013}
  M. S. Block, R. G. Melko, and R. K. Kaul,
  Phys. Rev. Lett. {\bf 111}, 137202 (2013).

\bibitem{PujariAD2015}  
  S. Pujari, F. Alet, and K. Damle,
  Phys. Rev. B {\bf 91}, 104411 (2015).

\bibitem{NahumSCOS2015}
  A. Nahum, J. T. Chalker, P. Serna, M. Ortu\~no, and A. M. Somoza,
  Phys. Rev. X {\bf 5}, 041048 (2015).

\bibitem{SreejithP2014}
  G. J. Sreejith and S. Powell,
  Phys. Rev. B {\bf 89}, 014404 (2014).

\bibitem{KuklovMPST2008}
  A. B. Kuklov, M. Matsumoto, N. V. Prokof’ev, B. V. Svistunov, and M. Troyer,
  Phys. Rev. Lett. {\bf 101}, 050405 (2008).

\bibitem{MotrunichV2008}
  O. I. Motrunich and A. Vishwanath,
  arXiv:0805.1494.

\bibitem{ShaoGS2016}
  H. Shao, W. Guo, and A. W. Sandvik, Science {\bf 352}, 213 (2016).

\bibitem{Nahum2015b}
  A. Nahum, P. Serna, J. T. Chalker, M. Ortu\~no, and A. M. Somoza,
  Phys. Rev. Lett. {\bf 115}, 267203 (2015).



\bibitem{Tang2013}
  Y. Tang and A. W. Sandvik, Phys. Rev. Lett {\bf 110}, 217213 (2013).

\bibitem{Senthil2006}
  T. Senthil and M. P. A. Fisher, Phys. Rev. B {\bf 74}, 064405 (2006)

\bibitem{SandvikE2010}  
  A. W. Sandvik and H. G. Evertz,
  Phys. Rev. B {\bf 82}, 024407 (2010).

\bibitem{LiangDA1988}
  S. Liang, B. Doucot, and P. W. Anderson,
  Phys. Rev. Lett. {\bf 61}, 365 (1988).

\bibitem{SandvikB2007}
  A. W. Sandvik and K. S. D. Beach, arXiv:0704.1469.

\bibitem{SuwaT2015}
  H. Suwa and S. Todo,
  Phys. Rev. Lett. {\bf 115}, 080601 (2015).

\bibitem{SenSS2015}
  A. Sen, H. Suwa, and A. W. Sandvik,
  Phys. Rev. B {\bf 92}, 195145 (2015).
  
 \bibitem{Nomura1995}
  K. Nomura,
  J. Phys. A: Math. Gen. {\bf 28}, 5451 (1995).

 \bibitem{Eggert1996}
  S. Eggert, Phys. Rev. B {\bf 54}, R9612 (1996).

 \bibitem{Sandvik2010c}
  A. W. Sandvik, Phys. Rev. Lett. {\bf 104}, 137204 (2010).
    
\bibitem{Bartosch2013}
  L. Bartosch,
  Phys. Rev. B {\bf 88}, 195140 (2013).

\bibitem{Singh1995}
  R. R. P. Singh and M. P. Gelfand, Phys. Rev. B {\bf 52}, R15695(R) (1995).
  
\bibitem{Sandvik2001}
  A. W. Sandvik and R. R. P. Singh, Phys. Rev Lett. {\bf 86}, 528 (2001).

\bibitem{Powalski2015}
  M. Powalski, G. S. Uhrig, and K. P. Schmidt,
  Phys. Rev. Lett. {\bf 115}, 207202 (2015).

\bibitem{SpanuBS2006}
  L. Spanu, F. Becca, and S. Sorella,
  Phys. Rev. B {\bf 73}, 134429 (2006).

\bibitem{Huh2013}
  Y. Huh, P. Strack, and S. Sachdev,
  Phys. Rev. Lett. {\bf 111}, 166401 (2013).
  
\bibitem{Ghioldi2016}
  E. A Ghioldi, M. G. Gonzalez, L. O. Manuel, and A. E. Trumper,
  Europhys. Lett. {\bf 113}, 57001 (2016).
  
\bibitem{Pollock1987}
  E. L. Pollock and D. M. Ceperley,
  Phys. Rev. B {\bf 36}, 8343 (1987).

  \bibitem{Kaul2008}
  R. K. Kaul and R. G. Melko,
  Phys. Rev. B {\bf 78}, 014417 (2008).

\bibitem{Jiang2011}
  F.-J. Jiang and U.-J. Wiese,
  Phys. Rev. B {\bf 83}, 155120 (2011); F.-J. Jiang, Phys. Rev. B {\bf 83}, 024419 (2011).
  
\end{thebibliography}
\end{document}